\documentclass[12pt]{article}

\makeatletter
\@addtoreset{equation}{section}

\renewcommand{\thefootnote}{\fnsymbol{footnote}}
\makeatother

\newcommand{\kahler}{K\"{a}hler }

\begin{document}

\thispagestyle{empty}
%
\begin{flushright}
TIT/HEP--470 \\
OU-HET 394 \\
{\tt hep-th/0108133} \\
August, 2001 \\
\end{flushright}
\vspace{3mm}
\begin{center}
{\Large
{\bf 
BPS Lumps and Their Intersections\\ 
in ${\cal N}=2$ SUSY Nonlinear Sigma Models
}}
\\[12mm]
\lineskip .75em
\vskip 1.3cm

\normalsize

  {\large \bf 
  Masashi~Naganuma~$^{a}$}
\footnote{\it  e-mail address: 
naganuma@th.phys.titech.ac.jp
},  
 {\large \bf 
Muneto~Nitta~$^{b}$}
\footnote{\it  e-mail address: 
nitta@het.phys.sci.osaka-u.ac.jp}\footnote{
Address after September 1, Department of Physics, West Lafayette, 
IN 47907-1396, USA.},
~and~~  {\large \bf 
Norisuke~Sakai~$^{a}$}
\footnote{\it  e-mail address: 
nsakai@th.phys.titech.ac.jp} 

\vskip 1.5em

{ \it $^{a}$Department of Physics, Tokyo Institute of 
Technology \\
Tokyo 152-8551, JAPAN  \\
and \\
  $^{b}$Department of Physics, Osaka University 
560-0043, JAPAN }
\vspace{18mm}
{\bf Abstract}\\[5mm]
{\parbox{13cm}{\hspace{5mm}
BPS lumps in 
${\cal N}=2$ SUSY nonlinear sigma models 
on hyper-\kahler manifolds in four dimensions are studied.   
We present new lump solutions with various kinds of 
topological charges.  
New BPS equations and a new BPS bound, 
expressed by the three complex structures on 
hyper-\kahler manifolds, are found. 
We show that any states satisfying these BPS equations 
preserve $1/8$ ($1/4$) SUSY of 
${\cal N}=2$ SUSY nonlinear sigma models 
with (without) a potential term. 
These BPS states  include non-parallel multi-(Q-)lumps.

}}
\end{center}
\vfill
\newpage
\setcounter{page}{1}
\setcounter{footnote}{0}
\renewcommand{\thefootnote}{\arabic{footnote}}

\section{Introduction}\label{INTRO}

Topological solitons saturating an energy bound, 
called the BPS bound~\cite{BPS,WittenOlive}, 
have played a crucial role in 
non-perturbative studies of supersymmetric 
(SUSY) field theories in four dimensions.  
Domain walls are topological solitons of co-dimension one, 
which depend on one spatial coordinate and connect two SUSY vacua.
Since they preserve half of the original SUSY in 
${\cal N}=1$ SUSY theories, 
they are called $1/2$ BPS states \cite{Gauntlett}.  
Such BPS domain walls were well studied 
in various models with 
${\cal N}=1$ SUSY~\cite{CGR,DvaliShifman}.  
The intersections or junctions of domain walls 
preserve $1/4$ SUSY in ${\cal N}=1$ 
models~\cite{AbrahamTownsend}--\cite{NNS}. 
For the case of ${\cal N}=2$ SUSY theories, 
the nontrivial interactions for hypermultiplets can exist only 
in nonlinear sigma models. 
Target spaces of ${\cal N}=2$ SUSY nonlinear sigma models 
must be hyper-\kahler (HK) manifolds~\cite{AF1}, 
and the only possible potential term is given by  
the square of a tri-holomorphic 
Killing vector acting on the HK manifold~\cite{AF2,Sierra}. 
These models are called ``massive HK nonlinear sigma models''.  
Single or parallel domain walls in such models 
preserve $1/2$ SUSY~\cite{AT,Townsend3}, 
whereas their intersections preserve $1/4$ SUSY~\cite{Townsend1}. 

On the other hand, there are 
topological solitons of co-dimension two, called lumps~\cite{Ward}. 
Lumps can exist in nonlinear sigma models 
whose target spaces are \kahler manifolds. 
By the dimensional reductions, 
along the direction of the lumps,  
to $(2+1)$- or $(2+0)$-dimensions, 
the lumps reduce to particle-like solitons~\cite{Ward} 
or instantons~\cite{BP}, respectively. 
The degree of the map from two-dimensional space to 
the \kahler manifold $M$ is expressed by 
the topological number, taking a value in 
the second homotopy class, $\pi_2(M)$~\cite{BP,Perelomov}. 
An instanton can exist when $\pi_2(M) \neq 0$, 
and moreover if $\pi_2(M) = {\bf Z}$, 
multi-instantons with the higher topological number can exist. 
The configuration of multi-instantons in two dimensions 
corresponds to the configuration of 
{\it parallel} multi-lumps in four dimensions. 
They preserve half of SUSY in the cases of 
SUSY nonlinear sigma models.  
The BPS bound is given by 
the pull-back of the \kahler form 
(a linear combination of 
the triplet of the \kahler forms) 
in \kahler (HK) manifolds to the real space.  
In the cases of the massive HK nonlinear sigma models, 
no static lumps can exist, 
but stationary lumps, called the ``Q-lumps'' 
can exist~\cite{Leese,Abraham}. 
They preserve $1/4$ SUSY and 
carry a Noether charge besides the topological charges \cite{Townsend2}. 
Recently, it has also been shown in \cite{Townsend2} 
that lumps can end on a domain wall in massive HK nonlinear sigma models 
in four dimensions, 
and this configuration is a $1/4$ BPS state. 
However, no $1/8$ BPS state has been found, 
although HK nonlinear sigma models have eight supercharges.  
These lumps can be considered a field theory realization of D-strings. 
Therefore we expect multi-lumps and/or intersections of lumps to exist. 
In this paper, we concentrate on the (Q-)lumps 
in (massive) HK nonlinear sigma models in four dimensions. 
As described above, they are $1/2$ BPS states 
and the BPS bound is given by 
the topological number represented by  
the pull-back of a linear combination of 
the triplet of \kahler forms on the HK manifolds~\cite{Ruback}. 
However, only known solution has been the lump in 
the Eguchi-Hanson gravitational instanton~\cite{EH}, 
which wraps around the two-cycle when interpolating two singularities of 
the manifold.  

The purpose of this paper is to work out the BPS equations suitable for the 
the $1/4$ ($1/8$) BPS states of non-parallel multi-lumps or 
intersections of lumps in the ${\cal N}=2$ SUSY massless (massive) HK 
nonlinear sigma models. 
We find new BPS equations and 
a new BPS bound given by 
the sum of three kinds of the pull-backs 
of the three \kahler forms of the HK manifold to 
the three {\it independent} planes in the real space.  
We also show that any solutions of these BPS equations 
preserve $1/4$ ($1/8$) SUSY in (massive) HK nonlinear sigma models.  
This is the first example of a $1/8$ BPS state 
in ${\cal N}=2$ SUSY field theories. 
Although we do not yet find concrete solutions for such configurations, 
we expect that these states are realized by the configurations 
of non-parallel (Q-)lumps or (Q-)lump intersections, because of the 
following four reasons. 
 First, our new BPS equations admit a parallel configuration of lumps 
 as a particular $1/2$ BPS solution. 
The solutions of our $1/4$ ($1/8$) BPS equations 
carry three independent topological charges, as 
required by the non-parallel multi-lump configurations.  
They preserve the {\it same} $1/4$ ($1/8$) 
SUSY as the non-parallel multi-(Q-)lump configurations. 
We can construct an explicit solution of our $1/4$ ($1/8$) BPS equations 
as multi-(Q-)lump intersections at least when we consider a direct 
product of ALE spaces where lumps are not interacting each other. 
These features suggests that our new BPS equations should contain 
 the non-parallel multi-(Q-)lump configurations as solutions.  
We find that non-parallel (Q-)lumps requires HK metric with 
at least three centers and 
give new lump solutions in multi-center 
asymptotically locally Euclidean 
(ALE) space of Gibbons-Hawking~\cite{GH}. 
These are not trivial extensions, since 
lump solutions connecting {\it different} pairs of centers 
preserve {\it different} combinations of the original SUSY. 
By analyzing the asymptotic behavior of multi-lump configurations, 
we find that the $1/4$ ($1/8$) SUSY requires that 
the direction of lumps in base space needs to be correlated with 
the choice of complex structures.

In sect.~2 we give a brief review of HK nonlinear sigma models 
and the general properties of the lump in these models. 
Then we work out the lump solutions 
in multi-center metric of HK target space, 
and discuss the relation between the possible configurations of lumps 
and centers in metric.
In sect.~3  we derive the BPS equation and BPS bound for $1/4$ ($1/8$) 
BPS state in $D=4$ (massive) HK nonlinear sigma models.
Moreover we discuss the SUSY condition for the non-parallel lumps, 
and we confirm that $1/4$ ($1/8$) SUSY is the same as the new state. 
Then we discuss that the new states can be expected to be the lump 
intersections.
Summary and discussions are given in sect.~4.

\vspace{5mm}
\section{1/2 BPS lumps in D=4 hyper-\kahler 
nonlinear sigma models}\label{sc:lump}
\subsection{Hyper-\kahler manifold}
The target space of the ${\cal N}=2$ 
SUSY nonlinear sigma models in four dimensions  
must be HK manifolds~\cite{AF1}.  
Let $\phi ^X$ $(X=1,2,..,4n)$ be the real 
coordinates of a $4n$-dimensional HK manifold, 
and ${{\rm I}^{(s)X}}_Y$ $(s=1,2,3)$ be 
the triplet of the complex structures.  
We use the vielbein formalism to describe a HK manifold. 
The 4n-beins $f^{ia}_X(\phi)$ have 
a pair of the tangent indices ($ia$) and 
transform as a $(2,2n)$-representation of $Sp_1 \times Sp_n$ 
under the local rotation in the tangent space 
of the HK manifold. 
The HK metric can be written in terms of $4n$-beins as 
\begin{eqnarray}
  g_{XY}(\phi) = f^{ia}_X(\phi) f^{jb}_Y(\phi) 
  \Omega _{ij}\epsilon _{ab}\;,
\end{eqnarray}
where $\Omega _{ij}$ and $\epsilon _{ab}$ are 
the real anti-symmetric tensors of $Sp_n$ and $Sp_1$, respectively. 
The inverse of the 4n-beins are $f_{ia}^X$: 
the equations $f^{ia}_X f_{ia}^Y = \delta_X^Y$ 
and $f^{ia}_X f_{jb}^X = \delta_j^i \delta_b^a$ hold. 
Another identity is also valid 
\begin{eqnarray}
f^{ia}_X f^Y_{ib} = \frac{1}{2}(\delta ^Y_X\delta ^a_b 
  - i\vec{\sigma} _b{}^a \cdot \vec{\rm I} ^Y{}_X) \;, 
\label{f-identity}
\end{eqnarray}
where $\vec{\sigma}$ is the triplet of Pauli matrices and 
$\vec{\rm I}$ is the triplet of complex structures. 
The triplet of \kahler forms associated 
with three complex structures, 
$\Omega^{(s)}_{XY} = g_{XZ}{I^{(s)Z}}_Y$, can be written as  
\begin{eqnarray}
 \vec{\Omega} = -\frac{i}{2}d\phi^X\wedge d\phi^Y f^{ia}_X f_{Yib}
                \vec{\sigma}_a{}^b \;. 
 \label{kahler-form}
\end{eqnarray}

\subsection{D=4 HK nonlinear sigma models}
One can obtain the $D = 4$ HK nonlinear sigma model
from $D=6$ HK nonlinear sigma model by a dimensional reduction. 
Let $\phi^X$ be the scalar fields parameterizing a HK manifold,  
and $\chi^i_{\alpha}$ be the fermionic partner. 
Here the index ${\alpha}$ transforms under $SU(4)^*$. 
It is known that there cannot exist any potential terms for  
$D=6$ hypermultiplets~\cite{Sierra}. 
The Lagrangian of the $D = 6$ HK nonlinear sigma model 
is given by \cite{Sierra} 
\begin{eqnarray}
{\cal L}=-\frac{1}{8}g_{XY}\partial^{\alpha \beta}\phi^X
      \partial_{\alpha \beta}\phi^Y
     -i\chi_{\alpha i} {\cal D}^{\alpha \beta} \bar{\chi}^i_\beta
     - \frac{1}{12}R_{ijkl}\chi^i_\alpha \chi^j_\beta 
       \chi^k_\gamma \chi^l_\delta 
       \varepsilon ^{\alpha \beta \gamma \delta}\;,
\end{eqnarray}
where $\varepsilon ^{\alpha \beta \gamma \delta}$ is 
the $SU(4)^*$ invariant tensor, 
and ${\cal D}^{\alpha \beta}$ is a covariant derivative. 
Here, we have taken the Minkowski metric 
as the `mostly plus' signature.
The SUSY transformation is given by
\begin{eqnarray}
 \delta \phi^X 
  &=& if^X_{ia}\epsilon ^{\alpha a}\chi ^i_\alpha \;,\\
 \delta \chi^i_\alpha 
  &=& f_X^{ia}\partial_{\alpha \beta}\phi^X \epsilon^\beta_a
     - \delta\phi^X\omega_{Xj}^i \chi^j_\alpha \;, 
\end{eqnarray}
where $\epsilon^\alpha_a$ is a constant spinor parameter 
of $Sp_1 \times SU(4)^*$. 

Let us derive the SUSY condition 
for a bosonic configuration, 
by requiring the SUSY transformation of the fermion to vanish.
Written in terms of the standard Dirac matrices, 
such a condition becomes
\begin{eqnarray}
  \Gamma^m f^{i a}_X \partial _m \phi^X \epsilon _a =0 \;,
  \label{susy-cond}
\end{eqnarray} 
where $\Gamma ^m$ ($m=0,1,...,5$) are 
the $D=6$ Dirac matrices, 
and $\epsilon_a$ is an $Sp_1$-Majorana and Weyl spinor, 
satisfying 
\begin{eqnarray}
  B \epsilon ^{*a} &=& \varepsilon ^{ab}\epsilon _b \;, \\
  \Gamma ^{012345}\epsilon &=& \epsilon \;.
\end{eqnarray} 
Here $B$ and $\Gamma^{012345}$ are given by 
$B=-i\rho^3\otimes \sigma^1\otimes \tau^2$ and 
$\Gamma^{012345}=\rho^3\otimes \sigma^3\otimes \tau^3$, 
in the standard representation of the Dirac matrices.
Then the spinor parameters $\epsilon_a$ of 
the SUSY transformation are expressed, 
using the four complex parameters $p,q,r,s$, by 
\begin{eqnarray}
  \epsilon_1 &=& (p,0,0,q,0,r,s,0) \;, \nonumber \\
  \epsilon_2 &=& (-q^*,0,0,p^*,0,-s^*,r^*,0) \;. 
\label{eq:susy_parameters}
\end{eqnarray}
Multiplying (\ref{susy-cond}) by $f^Y_{ib}$ and using the identity 
(\ref{f-identity}), the SUSY condition is equivalent to \cite{Townsend1}
\begin{eqnarray}
  \Gamma ^m \epsilon \partial _m\phi^Y g_{YX}
    -i\Gamma ^m \vec{\sigma}\epsilon
    \cdot \vec{\Omega}_{XY}\partial _m\phi^Y = 0 \;.
\label{susy-cond2}
\end{eqnarray} 

In four dimensions, 
massless HK nonlinear sigma models can be obtained by 
the trivial dimensional reduction:   
we obtain the SUSY condition by simply 
neglecting the terms labeled by $m=4,5$ in Eq.~(\ref{susy-cond2}).  
On the other hand, massive HK nonlinear sigma models 
can be obtained by the Scherk-Schwarz reduction~\cite{AF1,Sierra,Townsend1}, 
given by  
\begin{eqnarray}
  \partial_4\phi^X = \mu k^X(\phi), \quad \partial_5\phi^X = 0\;, 
 \label{SS-red.}
\end{eqnarray}
where $k^X(\phi)$ is a tri-holomorphic Killing vector of 
the HK manifold, and $\mu$ is a mass parameter. 
We have the scalar potential term  
\begin{eqnarray}
  V = \frac{\mu^2}{2}g_{XY}k^X (\phi) k^Y (\phi) \;,
 \label{HK-pot.}
\end{eqnarray}
in addition to the Lagrangian of the massless HK nonlinear sigma model. 
This type of the potential is the only possibility 
compatible with ${\cal N}=2$ SUSY~\cite{Townsend1}.
In this case, the SUSY condition can be obtained by 
substituting (\ref{SS-red.}) into the condition (\ref{susy-cond2}).

\subsection{Lumps and BPS 
condition}
The lump is a string-like topological soliton  
in $D=4$ HK nonlinear sigma models. 
First we derive the BPS equation for a static lump in 
massless HK nonlinear sigma models by minimizing 
the energy density ${\cal E}$. 
Choosing one complex structure ${\rm I}^{(s)}$ and the direction of 
the lump to be along $x^3$, we find  
\begin{eqnarray}
  {\cal E} &=& \frac{1}{2}g_{XY}(\dot{\phi^X}\dot{\phi^Y}
                + \partial_1\phi^X\partial_1\phi^Y 
                + \partial_2\phi^X\partial_2\phi^Y 
                + \partial_3\phi^X\partial_3\phi^Y) 
                \nonumber \\
           &=& \frac{1}{2}g_{XY}\dot{\phi^X}\dot{\phi^Y}
              +\frac{1}{2}g_{XY}\partial_3{\phi^X}\partial_3{\phi^Y}
                +\frac{1}{2}g_{XY}
                (\partial_1\phi^X \mp {{\rm I}^{(s)X}}_Z
                \partial_2\phi^Z)
                 (\partial_1\phi^Y \mp {{\rm I}^{(s)Y}}_W
                \partial_2\phi^W) \nonumber \\
           &{}& \pm {\Omega}^{(s)}_{XY}
                \partial_1\phi^X \partial_2\phi^Y\;,
\label{energydens}
\end{eqnarray}
where dots denote the differentiations with respect to 
the time coordinate. 
Therefore the BPS equations are 
$\partial_1\phi^X = \pm {{\rm I}^{(s)X}}_Y\partial_2\phi^Y, 
\; \partial_3\phi^X=0, \; \dot{\phi}^X = 0$.
If we denote the plane perpendicular to the direction $x^k$ of the lump 
in base space as $(x^i,x^j)$ and the invariant pseudo-tensor 
in the plane as $\varepsilon_{ij}$, the BPS equations are generalized to 
\begin{eqnarray}
  \partial_i\phi^X = \pm {{\rm I}^{(s)X}}_Y
   \partial_j\phi^Y, 
   \;\;\;\; 
   \partial_k\phi^X=0, 
   \;\;\;\; 
  \dot{\phi}^X = 0 \;,
\label{BPS-lump}
\end{eqnarray}
and the BPS bound becomes
\begin{eqnarray}
  L^{(s)} = \left|{1 \over 2} \int d^3x \Omega^{(s)}_{XY}\varepsilon_{ij} 
     \partial_i\phi^X\partial_j\phi^Y \right|\;. 
\end{eqnarray}
The integrand in the absolute value in right-hand-side of this equation 
is the pull-back of the \kahler form,  
$\Omega^{(s)}_{XY}=g_{XZ}{{\rm I}^{(s)Z}}_{Y}$, 
to the plane in the real space.

The BPS equations (\ref{BPS-lump}) satisfy the SUSY condition.
We can get these BPS equations by requiring the SUSY conservation 
(\ref{susy-cond2}) for the supercharge corresponding to spinor parameter 
satisfying 
\begin{eqnarray}
\Gamma ^{ij}\sigma ^{(s)}\epsilon = \pm i \epsilon \;, 
\label{1/2-lump}
\end{eqnarray}
Therefore, a static lump is a $1/2$ BPS state. 
Note that the condition (\ref{1/2-lump}) for supercharge  
corresponds to the lump solution which depends 
on the plane of ($x^i,x^j)$ and 
whose topological charge is the pull-back
of the \kahler form associated with 
the complex structure ${\rm I}^{(s)}$. 
Consider a lump which is extended along 
the direction of $(\theta,\phi)$ in the spherical coordinates of 
the real space, and has the topological 
charge associated with a linear combination of 
the three complex structures, 
$\alpha{\rm I}^{(1)}+\beta{\rm I}^{(2)}+\gamma{\rm I}^{(3)}$. 
Then, using a three vector $\vec{m}=(\alpha, \beta, \gamma)$, 
the SUSY condition of this lump is given by
\begin{eqnarray}
\tilde{\Gamma}(\vec{\sigma}\cdot \vec{m})\epsilon &=&
\tilde{\Gamma}(\alpha\sigma^{(1)}+\beta\sigma^{(2)}+\gamma\sigma^{(3)})
 \epsilon = \pm i\epsilon \;, \\
\tilde{\Gamma}&\equiv &\cos \theta \Gamma^{23}+\sin \theta\cos \phi 
\Gamma^{31}+\sin \theta \sin \phi \Gamma^{12} \; .
\label{general-gamma}
\end{eqnarray}
The topological charge of this lump becomes 
\begin{eqnarray}
 \alpha L^{(1)}+\beta L^{(2)}+\gamma L^{(3)}
 \equiv \vec{m}\cdot \vec{L}\; .
\end{eqnarray}
We thus have seen that $\vec{m}$ characterizes the linear combination 
 of the complex 
structure associated with topological charge of the lump.

Next we discuss the massive HK nonlinear sigma models. 
In these models, the potential term (\ref{HK-pot.}) 
should be added to the energy density (\ref{energydens}).  
We show that the BPS equations, 
the BPS bound and the conserved 
supercharges also change. 
The third equation of the BPS equations (\ref{BPS-lump}) 
is replaced by \cite{Abraham}
\begin{eqnarray}
  \dot{\phi}^X = \pm \mu k^X (\phi) \;,
\label{Q-lump}
\end{eqnarray}
using a tri-holomorphic Killing vector 
of the HK target manifold, $k^X (\phi)$.
The BPS bound is replaced to $L^{(s)}+Q$, 
where $Q$ is a conserved Noether charge, induced by 
the tri-holomorphic Killing vector:  
\begin{eqnarray}
  Q = \left| \mu \int d^3x g_{XY}\dot{\phi}^X k^Y(\phi) \right| \;.
  \label{eq:topological_charge}
\end{eqnarray}
In spite of the time-dependent configuration of this lump, 
the energy distribution is independent of the time.  
This situation is called stationary. 
This lump in massive HK nonlinear sigma models is called 
a ``Q-lump'' \cite{Abraham}.
Supercharges conserved by a Q-lump is 
given by Eq.~(\ref{1/2-lump}) and 
\begin{eqnarray}
 \Gamma^{04}\epsilon = \mp\epsilon \;,
\label{1/4-Qlump}
\end{eqnarray}
which corresponds to Eq.~(\ref{Q-lump}).
Therefore we find that 
a Q-lump is a $1/4$ BPS state from Eqs.~(\ref{1/2-lump}) and 
(\ref{1/4-Qlump}).

\subsection{Lump solution in multi-center metric}\label{sc:Multi}
\subsubsection{Multi-center models}
In this section, 
we shall consider a special class of SUSY nonlinear 
sigma models on a four-dimensional ALE space of the multi-center 
Gibbons-Hawking metric~\cite{GH}. 
Let the coordinates be $(\psi,\vec{X})$, 
with $\vec{X}$ being a three-vector, 
and $U(X)$ be a harmonic function. 
Then, the metric is given by 
\begin{eqnarray}
  ds^2 = Ud\vec{X}\cdot d\vec{X}+U^{-1}({\cal D}\psi)^2\;,
\end{eqnarray}
where ${\cal D}_k\psi\equiv \partial_k\psi 
+ \partial_k\vec{X}\cdot \vec{A}$ 
and $\vec{A}$ is related to $U$ by 
$\vec{\nabla}\times \vec{A}=\vec{\nabla}U$.
The triplet of the \kahler forms (\ref{kahler-form}) 
is expressed by 
\begin{eqnarray}
  \vec{\Omega}=(d\psi+d\vec{X}\cdot\vec{A})d\vec{X}
                -\frac{1}{2}Ud\vec{X}\times \vec{X} \;. 
\end{eqnarray}
We shall choose the harmonic function as 
\begin{eqnarray}
  U=\frac{1}{2}\sum^N_{i=1}\frac{1}{|\vec{X}-\vec{n}_i|} \;,
  \label{multi-U}
\end{eqnarray}
where $\vec{n}_i$ ($i=1,\cdots N$) are unit three-vectors. 
The function $U$ is singular at $\vec{X}=\vec{n}_i$, 
called the centers, 
but these are coordinate singularities of the metric
if $\psi$ is periodically identified with period $2\pi$. 
The ALE space can be considered as a $S^1$-bundle over 
the three vector space $\vec{X}$, 
with a fiber being parametrized by $\psi$. 
The isometry of $U(1)$ acts on the $S^1$ coordinate $\psi$,
and its action has fixed points at the centers.
Hence, a segment connecting each pair of two centers 
with $\psi$ parameterize a sphere $S^2$ as a submanifold. 
These spheres are called the $2$-cycles, 
and represent a nontrivial element of $\pi_2(M)$. 
For the function $U$ of Eq.~(\ref{multi-U}), 
we can set without loss of generality 
(see, e.g., \cite{NOY})
\begin{eqnarray}
 A^{(1)} 
  &=& \frac{1}{2}\sum_{i=1}^N \frac{X^{(2)}-n^{(2)}_i}
     {|\vec{X}-\vec{n}_i|(X^{(3)}-n^{(3)}_i-|\vec{X}-\vec{n}_i|)}
             \;, \nonumber \\
 A^{(2)} 
  &=& \frac{1}{2}\sum_{i=1}^N \frac{-(X^{(1)}-n^{(1)}_i)}
      {|\vec{X}-\vec{n}_i|(X^{(3)}-n^{(3)}_i-|\vec{X}-\vec{n}_i|)}
             \;, \nonumber \\
  A^{(3)} &=& 0 \;,
  \label{multi-A}
\end{eqnarray}
where we have chosen a gauge of $A^{(3)}=0$. 
In this multi-center models, SUSY condition of Eq.~(\ref{susy-cond2})
 becomes
\begin{eqnarray}
  [ \Gamma^m\vec{\sigma}\cdot \partial_m\vec{X}
     +iU^{-1}\Gamma^m{\cal D}_m\psi ] \epsilon = 0 \;. 
\label{SUSY-EH}
\end{eqnarray}

\subsubsection{General lump solution as holomorphic map}
Let us begin with the lump solution, 
extended along the $x^1$-axis 
(the configuration independent of $x^1$), 
and carrying the topological charge associated with complex 
structure ${\rm I}^{(1)}$.
This lump conserves $1/2$ SUSY for supercharges 
determined by 
\begin{eqnarray}
 \Gamma^{23}(\vec{\sigma}\cdot \vec{n})\epsilon = -i\epsilon, \quad 
 \mbox{for} \,\, \vec{n}=(1,0,0) \;. 
\end{eqnarray} 
~From the condition (\ref{SUSY-EH}) for this $1/2$ SUSY state, 
we get the BPS equations for the lump, given by  
\begin{eqnarray}
  \partial_k X^{(2)} &=& \partial_k X^{(3)} = 0 \quad 
  \mbox{for} \,\, k=1,2,3 \;,\nonumber \\
  \partial_1 X^{(1)} &=& {\cal D}_1 \psi = 0 \;,\nonumber \\
  {\cal D}_2\psi &=& U\partial_3X^{(1)} \;,\nonumber \\
  {\cal D}_3\psi &=& -U\partial_2X^{(1)} \;.
\label{1/2BPSeq}
\end{eqnarray}
~From the first and second of these equations, 
we find that $X^{(2)}=X^{(3)}= {\rm constant}$, 
and that $X^{(1)}$ and $\psi$ are independent of $x^1$.
Then the third and fourth can be combined together as
\begin{eqnarray}
(\partial_2+i\partial_3)[u-i(\psi+v)]=0 \;,
\end{eqnarray}
where we have defined real functions $u$ and $v$ by 
\begin{eqnarray}
 u(X^{(1)})\equiv \int dX^{(1)}U(X^{(1)}), \quad 
 v(X^{(1)})\equiv \int dX^{(1)}A^{(1)}(X^{(1)}) \;.
\label{uandv}
\end{eqnarray}
Therefore, the BPS equation for a lump 
can be reinterpreted as a {\it holomorphic map} from 
the complex plane $z \equiv x^2 + i x^3$ to the target space.  
The implicit form of the solution can be written as
\begin{eqnarray}
  \exp (u-i(\psi+v)) = Z(z) \;,
\end{eqnarray} 
where $Z(z)$ is a holomorphic function.
As a class of the solution, 
we can choose the holomorphic function as  
$Z=(z-z_0)^{-\alpha}$. 
Noting the reality of $u$, $v$ and $\psi$, 
we get 
\begin{eqnarray}
  X^{(1)} &=& u^{-1}\left( \frac{\alpha}{2}
   \log \frac{1}{|z-z_0|^2}\right) \;, \nonumber \\
  \psi &=& \alpha\cdot \arg (z-z_0) -v(X^{(1)}) \;.
  \label{general-sol}
\end{eqnarray}
Since $\psi$ is a periodic variable identification $\psi+2\pi=\psi$, 
the parameter $\alpha$ should be 
an integer, $\alpha \in {\bf Z}$. 
Therefore, we can interpret $\alpha$ as 
the topological number of the lump, 
 taking values in $\pi_2(M)$. 
Since $X^{(1)}$ depends only on 
the absolute value $|z-z_0|$, 
we can interpret the parameter $z_0$ as 
the position of the center of the lump in the $z$-plane. 
The BPS bound for this solution can be written 
in the polar coordinates as 
\begin{eqnarray}
  E \ge -\pi\alpha\int dx^1 \int dr \partial_r X^{(1)}
     = -\pi\alpha\int dx^1 [X^{(1)}(r=\infty)-X^{(1)}(r=0)], 
\end{eqnarray}
where $r\equiv \sqrt{(x^2-x_0^2)^2 + (x^3-x_0^3)^2}$. 
We see that the lump has an energy bound proportional to 
the topological number $\alpha$.

~From Eq.~(\ref{general-sol}), we see that $X^{(1)}$ approaches to 
$u^{-1}(\infty{\rm sign}(\alpha))$ 
in the limit of $r\to 0$, and to $u^{-1}(-\infty{\rm sign}(\alpha))$ 
in the limit of $r\to \infty$. 
Therefore both of $u^{-1}(\infty)$ and $u^{-1}(-\infty)$ must 
take finite values for the finiteness of 
the topological charge of the lump.

Next, we discuss the Q-lump in the massive HK nonlinear sigma models. 
As a tri-holomorphic Killing vector for 
the potential (\ref{HK-pot.}), 
we can choose the $U(1)$ isometry acting on $\psi$ as 
a constant shift.
Thus additional BPS equation for the Q-lump, 
Eq.~(\ref{Q-lump}), becomes 
\begin{eqnarray}
 \dot{\psi} = \pm \mu .
\end{eqnarray}
Therefore Q-lump solution can be obtained from 
the static lump-solution of Eq.~(\ref{general-sol}), 
replacing $\psi$ by 
\begin{eqnarray}
  \psi = \alpha\cdot \arg (z-z_0) -v(X^{(1)}) \pm \mu t \;.
\label{Qlump-sol}
\end{eqnarray}

\subsubsection{Lump in 2-center models}
Let us consider the simplest case of multi-center models, 
the 2-center metric. 
First, we set the two centers $\vec{n}_1=(1,0,0)$ and  
$\vec{n}_2=(-1,0,0)$. 
Setting $X^{(2)}=X^{(3)}=$ constant $=0$, 
the harmonic function becomes 
\begin{eqnarray}
  U=\frac{1}{2}\left( \frac{1}{|X^{(1)}-1|}+\frac{1}{X^{(1)}+1}\right)
   = \frac{1}{1-X^{(1)2}} \;, \quad A^{(1)} = 0 \;.
\end{eqnarray}
Then the equation $u(X^{(1)})=\tanh ^{-1} (X^{(1)})$ holds,  
and the solution of the lump 
can be obtained as \cite{Townsend2},
\begin{eqnarray}
  X^{(1)} &=& \tanh \left( \frac{\alpha}{2}\log \frac{1}{|z-z_0|^2}
              \right) \;, \nonumber \\
  \psi &=& \alpha\cdot \arg (z-z_0)\;.
\label{2center-sol}
\end{eqnarray}
Two limits of $X^{(1)}$ correspond to the positions of 
two centers, and therefore finite: $X^{(1)}=1$ for $r\to 0$, 
and $X^{(1)}=-1$ for $r\to \infty$. 
Hence, the topological charge of the lump is finite.
We note that the parameter $\alpha$ determines the size of lump 
as well as the topological number: 
we can see from Eq.~(\ref{2center-sol}) that 
the size of lump is greater 
as the topological number gets larger.  

It is important to realize that the finiteness of the topological charge 
imposes a severe constraint on the direction of the lump in base space 
$x^1, x^2, x^3$ and the direction of the space-dependent component of 
the field $\vec{X}$. 
To illustrate the point, let us consider, 
for example, a lump  solution in the case of two centers placed at 
$\vec{n}_1=(0,1,0)$ and  $\vec{n}_2=(0,-1,0)$.  
The BPS equation (\ref{1/2BPSeq}) implies that a lump can be constructed 
only by choosing $X^{(1)}$ space-dependent. 
In this case, the space-dependent component of $\vec{X}$ takes 
the {\it different} direction in the field space from the vector 
connecting the two centers in the metric. 
Setting $X^{(2)}=X^{(3)}=0$, 
we obtain 
\begin{eqnarray}
 U = \log (\sqrt{(X^{(1)})^2+1}+X^{(1)}), \quad A^{(1)}=0 \;. 
\end{eqnarray}
The implicit form of the solution of lump can be 
obtained as 
\begin{eqnarray}
\frac{\alpha}{2}\log \frac{1}{|z-z_0|^2}
&=& 
\log (\sqrt{(X^{(1)})^2+1}+X^{(1)}) \;,\nonumber \\
\psi &=& \alpha\cdot \arg (|z-z_0|) \;. 
\end{eqnarray}
However $X^{(1)}$ does not take finite values 
in the limit of both $r\to 0$ and $r\to \infty$. 
Therefore the solution gives a divergent topological charge and is 
unacceptable. 
To obtain a finite topological charge, it is necessary to align the 
direction of the lump in the base space with the direction connecting 
two centers of the metric.

We can understand this situation from the topological point of view. 
A lump is an axially-symmetric soliton, 
and the center of the lump ($r=0$) and 
the infinity around the lump ($r=\infty$) are mapped 
to two different points in the target space $M$, 
respectively.  
When these two points coincide with 
two different centers in the target space respectively as 
in the case of Eq.~(2.39), the map is closed  
and the lump is wrapped around a $2$-cycle. 
On the other hand, when any one of these two points are not mapped to 
a center as in the case of Eq.~(2.41), 
the map is not closed. 
We have shown that if and only if the map is closed, 
the the lump charge is finite, and takes a value in $\pi_2(M)$.  
A similar requirement on the finiteness of the topological charge 
was previously considered before \cite{DLD,Abraham}. 

\subsubsection{Lumps in 3-center models}
We see that the direction connecting the two centers 
must be aligned with  the direction of the complex structure 
characterizing the lump. 
Hence, to study the non-parallel lumps or lump intersections,  
we need to consider the multi-center model 
where centers are not aligned. 
Let us consider 3-center models of the four-dimensional HK metric as 
the simplest example.

As an example, we can work out an implicit solution of  
lumps in the case of the metric with three centers st 
$\vec{n}_1=(0,1,0)$, $\vec{n}_2 =(1,0,0)$ and $\vec{n}_3=(-1,0,0)$.
We consider two kinds of lumps in this model: 
The first lump corresponds to 
the line segment connecting the two centers at 
$\vec{n}_1$ and $\vec{n}_2$ in the 
field space $\vec{X}$, 
and the second lump corresponds to the two centers at 
$\vec{n}_1$ and $\vec{n}_3$. 

Now we find the solution of the first lump. 
For this purpose, we transform
the variable $\vec{X}$ to $\vec{Y}$ by  
\begin{eqnarray}
  \left(
  \begin{array}{c}
    Y^{(1)} \\
    Y^{(2)}
  \end{array}
  \right) \equiv \frac{1}{\sqrt{2}}
  \left(
  \begin{array}{c}
    X^{(1)}-X^{(2)}+1 \\
    X^{(1)}+X^{(2)}-1
  \end{array} 
  \right) \;. 
\end{eqnarray} 
In the basis of $\vec{Y}$, lines connecting two centers $\vec{n}_1$ and 
$\vec{n}_2$ are along the $Y^{(1)}$ axis in the field space. 
Therefore $Y^{(1)}$ becomes 
space-dependent for the first lump. 
Setting $Y^{(2)}=Y^{(3)}=0$, we have 
\begin{eqnarray}
  U &=& \frac{1}{2}\left[ \frac{1}{\sqrt{(Y^{(1)})^2}} 
         + \frac{1}{\sqrt{(Y^{(1)}-\sqrt{2})^2}} 
         +\frac{1}{\sqrt{(Y^{(1)})^2+2}} \right] \nonumber \\
    &=&  \frac{1}{2}\left[ \frac{2\sqrt{2}}{1-(\sqrt{2}Y^{(1)}-1)^2} 
         + \frac{1}{\sqrt{(Y^{(1)})^2+2}} \right] \;, \nonumber \\ 
  A^{(1)} &=& -\frac{1}{2}\frac{\sqrt{2}}{(Y^{(1)})^2+2} \;.  
\end{eqnarray}
Then the functions $u$ and $v$ are given from Eq.~(\ref{uandv}) by 
\begin{eqnarray}
  u = \tanh ^{-1}(\sqrt{2}Y^{(1)}-1)+\frac{1}{2}\log (Y^{(1)}
          +\sqrt{(Y^{(1)})^2+2}),   \quad 
  v = -\frac{1}{2} \tan ^{-1}
      \left( \frac{Y^{(1)}}{\sqrt{2}} \right) \;. 
\end{eqnarray}
We can find the implicit solution of the first lump 
from Eq.~(\ref{general-sol}):
\begin{eqnarray}
  -\frac{\alpha}{2}\log |z-z_0|^2 &=& 
    \tanh ^{-1}(X^{(1)}-X^{(2)}) \nonumber \\
    &{}& + \frac{1}{2}\log \left(
          \frac{1}{\sqrt{2}}(X^{(1)}-X^{(2)}+1)
          +\sqrt{\frac{1}{2}(X^{(1)}-X^{(2)}+1)^2+2}
          \right) \;, \nonumber \\
  \psi &=& \alpha\cdot \arg (z-z_0) +\frac{1}{2}\tan ^{-1}\left(  
    \frac{1}{2}(X^{(1)}+X^{(2)}-1)
    \right) 
    \label{lump-1st}
\end{eqnarray}
where $z\equiv x^2+ix^3$. 
The second lump is obtained similarly to give 
\begin{eqnarray}
  -\frac{\alpha}{2}\log |z-z_0|^2 &=& 
    \tanh ^{-1}(X^{(1)}+X^{(2)}) \nonumber \\
    &{}& + \frac{1}{2}\log \left(
         \frac{1}{\sqrt{2}}(X^{(1)}+X^{(2)}-1)
         +\sqrt{\frac{1}{2}(X^{(1)}+X^{(2)}-1)^2+2}
         \right) \;, \nonumber \\
  \psi &=& \alpha\cdot \arg (z-z_0) +\frac{1}{2}\tan ^{-1}\left(  
    \frac{1}{2}(X^{(1)}-X^{(2)}+1)
    \right) .
\label{lump-2nd} 
\end{eqnarray}

We can extend these solutions 
to lumps oriented to other directions in 
the multi-center model 
straightforwardly. 

\vspace{5mm}

\section{BPS equation for coexisting lumps}\label{sc:BPSeq}
In the last section, 
we have found the several kinds of lumps with different 
topological charges. 
Each (Q-)lump conserves different $1/2$ ($1/4$) SUSY in massless (massive) 
HK nonlinear sigma models.   
In this section, we derive new BPS equations and BPS bound 
in $D=4$ HK nonlinear sigma models. 
Any solutions of these BPS equations conserve $1/4$ SUSY 
(or $1/8$ SUSY in massive HK nonlinear sigma models).
We also derive conditions for 
coexisting lumps to preserve 
$1/4$ SUSY (or $1/8$ in massive theory). 

\subsection{New BPS bound and BPS equation in HK nonlinear sigma models}
The previous $1/2$ BPS equation (\ref{BPS-lump}) used only one complex 
structure. 
Since there are three complex structures in the HK manifolds, it should be 
possible to use all three complex structures to derive a new BPS bound. 
 For that purpose, we consider a configuration in $D=4$ HK nonlinear 
 sigma models which depends on three 
independent coordinates $x^1, x^2, x^3$ in base space. 
The energy density of this solitonic configuration can be rewritten as 
\begin{eqnarray}
  {\cal E} &=& \frac{1}{2}g_{XY}(\dot{\phi^X}\dot{\phi^Y}
                + \partial_1\phi^X\partial_1\phi^Y 
                + \partial_2\phi^X\partial_2\phi^Y 
                + \partial_3\phi^X\partial_3\phi^Y) 
                \nonumber \\
        &=& \frac{1}{2}g_{XY}\dot{\phi}^X\dot{\phi}^Y \nonumber \\
        &{}&   + \frac{1}{2}g_{XY}
               (\partial_1\phi^X - {\tilde{\rm I}}^{(2)X}{}_Z \partial_3\phi^Z
                 - {\tilde{\rm I}}^{(3)X}{}_Z \partial_2\phi^Z)
               (\partial_1\phi^Y - {\tilde{\rm I}}^{(2)Y}{}_W \partial_3\phi^W
                 - {\tilde{\rm I}}^{(3)Y}{}_W \partial_2\phi^W) \nonumber \\
        &{}&   +\Omega^{(1)}_{XY}\partial_2\phi^Y\partial_3\phi^X 
               + \tilde{\Omega}^{(2)}_{XY}\partial_3\phi^Y\partial_1\phi^X 
               + \tilde{\Omega}^{(3)}_{XY}\partial_2\phi^Y\partial_1\phi^X \nonumber \\      
        &\ge & \Omega^{(1)}_{XY}\partial_2\phi^Y\partial_3\phi^X 
               + \tilde{\Omega}^{(2)}_{XY}\partial_3\phi^Y\partial_1\phi^X 
               + \tilde{\Omega}^{(3)}_{XY}\partial_2\phi^Y\partial_1\phi^X\;. 
           \label{BPSbound}
\end{eqnarray}
where rotated complex structures $\tilde{\rm I}^{(2)}\equiv \vec{n}_2\cdot \vec{\rm I},\;
\tilde{\rm I}^{(3)}\equiv \vec{n}_3\cdot \vec{\rm I}$, and 
$\vec{n}_2\equiv (0, \cos\omega, \sin\omega),\;
\vec{n}_3\equiv (0, -\sin\omega, \cos\omega)$, are defined with a parameter $\omega$. 
This BPS inequality is useful 
for the solitonic configuration which depends on three coordinates. 
Moreover, this inequality is saturated when the configuration satisfies 
the following BPS equations 
\begin{eqnarray}
\partial_1\phi^X = {\tilde{\rm I}}^{(2)X}{}_Y \partial_3\phi^Y
                 + {\tilde{\rm I}}^{(3)X}{}_Y \partial_2\phi^Y
\; , \quad \dot{\phi}^X=0 \; .
  \label{BPSeq-int}
\end{eqnarray} 
In this case, 
the BPS bound for the soliton in Eq.~(\ref{BPSbound}) is written as 
a sum of three kinds of the pull-backs of \kahler forms 
to three planes in the real space, 
representing three topological charges. 

Next we show that $1/4$ SUSY remains unbroken for the solution 
of Eq.~(\ref{BPSeq-int}).
By substituting  Eq.~(\ref{BPSeq-int}) to the SUSY condition 
(\ref{susy-cond2}), we obtain 
\begin{eqnarray}
0 &=& {\rm I}^{(1)X}{}_Y\partial_2\phi^Y[\Gamma^3 - i\Gamma^2\sigma^{(1)}]\epsilon
     + {\rm I}^{(1)X}{}_Y\partial_3\phi^Y[-\Gamma^2 - i\Gamma^3\sigma^{(1)}]\epsilon 
       \nonumber \\
 &{}& +\tilde{\rm I}^{(2)X}{}_Y\partial_3\phi^Y[\Gamma^1 - i\Gamma^3
      (\vec{\sigma}\cdot \vec{n}_2)]\epsilon
     + \tilde{\rm I}^{(2)X}{}_Y\partial_1\phi^Y[-\Gamma^3 - i\Gamma^1
       (\vec{\sigma}\cdot \vec{n}_2)]\epsilon 
       \nonumber \\
 &{}& +\tilde{\rm I}^{(3)X}{}_Y\partial_2\phi^Y[\Gamma^1 - i\Gamma^2
       (\vec{\sigma}\cdot \vec{n}_3)]\epsilon
     + \tilde{\rm I}^{(3)X}{}_Y\partial_1\phi^Y[-\Gamma^2 - i\Gamma^1
       (\vec{\sigma}\cdot \vec{n}_3)]\epsilon 
       \nonumber \\
 &{}& -i{\rm I}^{(1)X}{}_Y\partial_1\phi^Y[\Gamma^1\sigma^{(1)} 
        + \Gamma^3(\vec{\sigma}\cdot \vec{n}_3)]\epsilon 
      -i\tilde{\rm I}^{(2)X}{}_Y\partial_2\phi^Y[\Gamma^2(\vec{\sigma}\cdot \vec{n}_2) 
        + \Gamma^3(\vec{\sigma}\cdot \vec{n}_3)]\epsilon 
   \;.
\end{eqnarray}
The first and the third terms should vanish, giving 
\begin{eqnarray}
  \Gamma^{23}\sigma^{(1)}\epsilon  
    = -i\epsilon \; , \quad
  \Gamma^{31}(\vec{\sigma}\cdot \vec{n}_2)\epsilon 
    = -i\epsilon. 
\label{SUSY-int1} 
\end{eqnarray}
These two conditions are sufficient to make all the other terms vanish. 
By calculating Eq.~(\ref{SUSY-int1}) for Dirac matrices in the 
standard representation, the conserved supercharges are given by 
the following relations for the components of the spinor parameters 
in Eq.~(\ref{eq:susy_parameters}) 
\begin{eqnarray}
  q=-p^*, \,\, r=-s^* \,\, \mbox{and} \,\, p=-e^{-i\omega}s \;. 
\label{1/4SUSY}
\end{eqnarray}
~From these relations, we see that $1/4$ SUSY remains unbroken. 

In the case of massive HK nonlinear sigma models, 
this soliton can carry the Noether charge (\ref{eq:topological_charge}) 
induced by the isometry 
of the target space, in addition to the three topological terms of 
the BPS bound in Eq.~(\ref{BPSbound}). 
Correspondingly, 
the second equation of Eq.~(\ref{BPSeq-int}) is replaced 
by Eq.~(\ref{Q-lump}), and the condition (\ref{1/4-Qlump}) 
is required as additional constraint for conserved supercharges. 
Eq.~(\ref{1/4-Qlump}) is rewritten, in terms of the components of 
the spinor parameters, as 
\begin{eqnarray} 
 p = \mp i s^* .
\label{1/8SUSY}
\end{eqnarray}
Eqs.~(\ref{1/4SUSY}) and (\ref{1/8SUSY}) together imply that there is only 
one real parameter for the spinor transformation parameters. 
Therefore 
this soliton is $1/8$ BPS state and conserves minimal SUSY 
in ${\cal N}=2$ four-dimensional massive HK nonlinear sigma models. 

We expect that the solutions of the BPS equation Eq.~(\ref{BPSeq-int}) 
contain the intersecting or coexisting lumps for several reasons. 
The first reason is 
that the BPS bound is given by the sum of three kinds of the 
pull-backs of the \kahler forms to three planes in the real space. 
Each pull-back corresponds to the topological charge of the lump. 
The second is that the BPS equation of a single lump can be derived as 
the special case of Eq.~(\ref{BPSeq-int}).
If we require the fields to be independent of the $x^1$, for example, 
we can recover, from Eq.~(\ref{BPSeq-int}), the BPS equation of the 
lump such as 
\begin{eqnarray}
  \partial_1\phi^X=0\;,\quad 
  \partial_2\phi^X= - {{\rm I}^{(1)X}}_Y\partial_3\phi^Y.
\end{eqnarray}
As the third reason, we confirm, in the next section, that the states 
for coexisting or intersecting lumps can conserve the {\it same}
\footnote{
Intersecting domain walls considered in Ref.\cite{Townsend1} conserve 
{\it different} $1/4$ SUSY.
} 
 $1/4$ SUSY ($1/8$ SUSY in massive theory) as the new soliton 
 that we discussed above.
The fourth reason is that 
we can construct an explicit solution of 
multi-lump intersections as a $1/4$ BPS state
at least when we consider the direct product of ALE spaces, 
as discussed below.

\subsection{BPS condition for coexistence of lumps}
Let us consider generally the BPS condition for coexistence of two lumps here. 
For simplicity, the first lump is extended along the $x^1$-direction 
and is associated with the complex structure ${\rm I}^{(1)}$. 
The SUSY condition for this lump can be written as  
\begin{eqnarray}
\Gamma^{23}(\vec{\sigma}\cdot \vec{n}) \epsilon 
 = -i\epsilon \quad 
\mbox{for} \quad \vec{n}=(1,0,0) \;. 
\label{1st-lump}
\end{eqnarray}
Next we consider the second lump and look for the condition for the 
second lump to retain a part of SUSY preserved by the first one. 
We try to put the second lump in the direction of $(\theta, \phi)$ 
in the spherical coordinates.
Hence, SUSY condition for the second lump is 
\begin{eqnarray}
\tilde{\Gamma}(\vec{\sigma}\cdot \vec{m}) \epsilon = -i\epsilon 
\quad \mbox{for} \quad \vec{m}
=(\cos\theta_c, \sin\theta_c\cos\phi_c, \sin\theta_c\sin\phi_c)\;, 
\label{2nd-lump}
\end{eqnarray}
where 
$\tilde{\Gamma}$ is the linear combination of the Dirac matrices 
given in Eq.~(\ref{general-gamma}). 
Since the coexistence of the two lumps requires 
 the spinor parameter to satisfy 
the two constraints (\ref{1st-lump}) and (\ref{2nd-lump}). 
By demanding partial conservation of SUSY, we find that 
the complex structure has to be correlated with the lump direction:
$\theta_c = \theta $, 
and that the conserved SUSY is given in terms of spinor components 
in (\ref{eq:susy_parameters}) as 
\begin{eqnarray}
  q=-p^*, \,\, r=-s^* \,\, \mbox{and} \,\, p=-e^{-i(\phi+\phi_c)}s,  
\end{eqnarray}
in the standard representation of Dirac matrices.
This equation corresponds to Eq.(\ref{1/4SUSY}), by setting 
$\omega=\phi+\phi_c$. 
Therefore non-parallel multi-lump configurations with the 
complex structure satisfying the relation $\theta = \theta_c$ 
preserves $1/4$ SUSY. 
In other words, a $1/4$ SUSY is conserved 
when several lumps coexist or intersect with arbitrary angles, 
if the lump extended along the direction of $(\theta, \phi)$   
carries the topological charge corresponding to 
\begin{eqnarray}
\vec{m}=(\cos\theta, \sin\theta\cos(\omega-\phi), \sin\theta\sin(\omega-\phi)),
\label{correlation2}
\end{eqnarray} 
for a fixed parameter $\omega$.

We note that the Eq.~(\ref{BPSeq-int}) can allow the many solutions for 
coexisting lumps which are extended along the arbitrary directions 
in the real space, if the lumps pick up the particular complex structure 
as in Eq.~(\ref{correlation2}):
We can see from Eq.~(\ref{2nd-lump}) that there is a two (real) parameter 
family $(\theta, \phi)$ of lump 
configurations conserving the same supercharges 
as given in Eq.~(\ref{1/4SUSY}). 
In the case of coexisting Q-lumps, 
$1/8$ SUSY corresponding to Eqs.(\ref{1/4SUSY}) and (\ref{1/8SUSY})
is conserved if each of the coexisting (intersecting) Q-lumps picks up the 
particular complex structure given in Eq.~(\ref{correlation2}). 
Hence the coexisting Q-lumps conserve minimal SUSY 
in ${\cal N}=2$ four-dimensional massive HK nonlinear sigma models. 

\subsection{Lump intersections and multi-center metric}
We can see from Eq.~(\ref{correlation2}) that, if non-parallel lumps 
coexist, such lumps must have 
different topological charges. 
But lumps with various topological charges cannot always exist in 
general HK nonlinear sigma models. 
We wish to consider the situation where several lumps with different 
topological charges can coexist. 
In the model with multi-center metric, several centers 
should not be placed on a single line 
in field space, since several lumps should wrap topologically 
different 2-cycles of the target space 
oriented along different directions. 
Conversely, 
possible directions of different lumps are determined by the positions 
of centers for a given model. 
We think that the 3-center model in previous section is a candidate 
of the simplest model of four-dimensional HK metric 
with possibility of two lumps coexisting 
perpendicularly.

An explicit solution of 
a $1/4$ ($1/8$) BPS state 
of massless (massive) nonlinear sigma models 
can be realized as follows. 
First, we prepare a $4n$-dimensional target space as  
a direct product of $n$ ALE spaces. 
In this case, fields are divided into $n$ sectors, 
which are not interacting  each other. 
This model admits a solution of $n$-lump intersection, 
in which each lump has proper 
correlation of (\ref{correlation2}). 
This provides an explicit example of a $1/4$ ($1/8$) BPS state 
in massless (massive) nonlinear sigma models, 
although each lump is not interacting with the others.
By introducing a constant in 
the off-diagonal block of the product metric, 
we may construct 
a more interesting example of 
interacting lumps, 
as was done in a wall intersection~\cite{Townsend1}.

\vspace{5mm}

\section{Summary and discussions}
\label{sc:Summary}
We have examined several kinds of (Q-)lumps with various topological 
charges in $D=4$ hyper-\kahler SUSY nonlinear sigma models. 
Especially 
in the model 
with multi-center metric, we have worked 
out the implicit solutions for a general 
lump. 
In order to obtain a lump with finite energy density, 
the field $\vec{X}$ has to be aligned along the direction 
connecting two centers. 
Otherwise, the configuration does not wrap the $2$-cycle connecting 
the two centers and the topological charge diverges. 

Next, we have found the BPS equation and energy bound for a soliton 
which depends on three spatial coordinates 
 in general HK nonlinear sigma models. 
The BPS bound is written as a sum of three kinds of the pull-backs 
of \kahler forms to three planes in the real space. 
We also show that solutions of the BPS equation corresponds to 
$1/4$ BPS states, and the BPS equation admits a single lump solution 
preserving $1/2$ SUSY as a special case. 
In massive HK nonlinear sigma models, the soliton conserves $1/8$ SUSY. 

Moreover, we also have considered the SUSY conditions for general 
coexisting or intersecting lumps. We found that 
$1/4$ SUSY is conserved by coexisting lumps and $1/8$ SUSY is conserved 
by coexisting Q-lumps, when the directions of lumps correlate properly 
to the topological charges carried by the lumps. 
Hence, we expect that new soliton corresponds to the intersecting or 
coexisting lumps. 
Partial conservation of SUSY requires that the polar angle $\theta_c$ 
in the space of complex structures has to be the same 
as the polar angle $\theta $ in the real space for the second lump 
relative to the first. 
The sum of the azimuthal angle $\phi_c$ in the space of complex 
structures and the azimuthal angle $\phi$ determines 
the conserved SUSY.
Therefore we can consider $1/4$ SUSY conservation of arbitrarily many 
lumps provided the above condition is met with a fixed 
$\omega = \phi_c + \phi$.

To find an analytic solution of coexisting lumps with different 
topological charges is a nontrivial problem. 
When we consider coexisting lumps in multi-center models of four-dimensional 
HK metric, the directions of lumps should be related to the directions 
of lines connecting pairs of centers in the multi-center metric. 
It is very interesting whether the solitons 
wrapped in different 2-cycles can coexist. 

It has been found that lumps can end on the domain wall 
and that this configuration preserves $1/4$ SUSY~\cite{Townsend2}. 
In Ref.~\cite{Townsend2}, it has been argued that much of the physics of 
D-branes can appear as Q-lumps in a purely field theoretical context, 
in particular a D-string ending on a D-brane. 
Our results on the non-parallel or intersecting lumps will shed more light on 
the study of D-branes in a field theoretical context. 

\vspace{.5cm}
{\bf Acknowledgement}:
This work is supported in part by Grant-in-Aid 
for Scientific Research from the Japan Ministry 
of Education, Science and Culture for 
the Priority Area 707 and 13640269. 
We also thank the Yukawa Institute for Theoretical Physics 
at Kyoto University.  
Discussions during the YITP workshop 
on "Quantum Field Theory 2001"  were useful to complete this work. 



\end{document}